\documentclass[aps,prl,twocolumn,amsmath,amssymb,superscriptaddress]{revtex4-2}

\usepackage{amsmath}
\usepackage{graphicx}
\usepackage{dcolumn}
\usepackage{bm}
\usepackage{multirow}
\usepackage{subfigure}
\usepackage{booktabs}
\usepackage{hhline}  
\usepackage{float}
\usepackage{braket}
\usepackage[colorlinks,linkcolor=blue,anchorcolor=blue,citecolor=blue,urlcolor=blue]{hyperref}
\usepackage[normalem]{ulem}
\usepackage{mathtools}
\usepackage{xcolor}
\usepackage{mathbbol}
\usepackage{yhmath}
\usepackage{todonotes}
\usepackage{lineno}

\begin{document}
	
	
	\title{Nonadiabatic Theory of Phonon Magnetic Moments in Insulators and Metals}
	
	
	\author{Haoran Chen}
	\affiliation{Department of Materials Science and Engineering, University of Washington, Seattle 98195, United States}
	
	\author{Wenqin Chen}
	\affiliation{Department of Physics, University of Washington, Seattle, Washington 98195, United States}
	
	\author{Kaijie Yang}
	\affiliation{Department of Materials Science and Engineering, University of Washington, Seattle 98195, United States}
	
	\author{Ting Cao}
	\email{tingcao@uw.edu}
	\affiliation{Department of Materials Science and Engineering, University of Washington, Seattle 98195, United States}
	
	\author{Di Xiao}
	\email{dixiao@uw.edu}
	\affiliation{Department of Materials Science and Engineering, University of Washington, Seattle 98195, United States}
	\affiliation{Department of Physics, University of Washington, Seattle, Washington 98195, United States}
	
	
	\date{\today}
	
	\begin{abstract}
    We develop a nonadiabatic theory of phonon magnetic moments applicable to both insulators and metals. By relating the phonon magnetic moment to the force-velocity response of ions in a magnetic field, we derive a gauge-invariant expression using a gauge-covariant Wigner expansion. The formalism naturally separates Fermi-sea and Fermi-surface contributions and captures the full dependence on phonon frequency. In gapped systems, our theory reduces to previous adiabatic expressions in the low-frequency limit. Beyond this limit, it reveals additional contributions arising from resonant interband processes and the Fermi surface. Applying our theory to Pb$_{1-x}$Sn$_x$Te, we find that the Fermi-surface contribution substantially enhances the phonon magnetic moment, reproducing the same order of magnitude as the experimental observation.
    Our results provide a unified framework for describing phonon magnetic moments beyond the adiabatic regime.
	\end{abstract}
	\maketitle
	
	\textit{Introduction.}---Phonons are among the most fundamental quasiparticles in condensed matter, playing a central role in thermodynamic, spectral, and transport properties of materials~\cite{giustino_electron-phonon_2017}.
	In the past two decades, phonons carrying finite angular momentum, known as chiral or axial phonons, have attracted growing interest due to their nontrivial modification of these properties~\cite{qin_berry_2012,zhang_angular_2014,juraschek_dynamical_2017,zhu_observation_2018,xu_nondegenerate_2018,juraschek_orbital_2019,saparov_lattice_2022,juraschek_giant_2022,ueda_chiral_2023,juraschek_chiral_2025,shabala_axial_2025,bonini_frequency_2023,ren_phonon_2021,zhang_advances_2025,yang_inherent_2025,che_magnetic_2025,li_phonon_2025}. 
	They are closely related to various intriguing phenomena such as the phonon Zeeman effect~\cite{juraschek_dynamical_2017,juraschek_orbital_2019,juraschek_giant_2022,cheng_large_2020,baydin_magnetic_2022,hernandez_observation_2023}, the thermal Hall effect~\cite{strohm_phenomenological_2005,qin_berry_2012,ideue_giant_2017,saito_berry_2019,zhang_thermal_2019,boulanger_thermal_2020,li_phonon_2020,chen_enhanced_2020,sun_phonon_2021,zabalo_rotational_2022,li_phonon_2023,ataei_phonon_2024,oh_phonon_2025,mangeolle_extrinsic_2026}, and the Einstein-de Haas effect~\cite{zhang_angular_2014,dornes_ultrafast_2019,zhang_measurement_2025}. 
	Moreover, due to their nature of broken time-reversal symmetry, chiral phonons also carry magnetic moments, allowing the exciting possibility of generating spin and orbital magnetizations dynamically~\cite{nova_effective_2017,holanda_detecting_2018,disa_polarizing_2020,afanasiev_ultrafast_2021,stupakiewicz_ultrafast_2021,mashkovich_terahertz_2021,sasaki_magnetization_2021,luo_large_2023,ren_light-driven_2024,basini_terahertz_2024,shabala_phonon_2024,klebl_ultrafast_2025}.
	
    Despite substantial theoretical progress, a quantitative understanding of phonon magnetic moments remains incomplete. Existing quantum theories, such as those based on adiabatic current pumping~\cite{zhang_gate-tunable_2023,xiao_adiabatically_2021,ren_phonon_2021,trifunovic_geometric_2019}, density functional perturbation theory~\cite{zabalo_rotational_2022}, and density-matrix perturbation approaches~\cite{xue_extrinsic_2025}, are primarily formulated for insulating systems. However, recent experiments in doped semiconductors and metals have reported phonon magnetic moments reaching the scale of $\mu_B$, far exceeding the predictions of these approaches~\cite{cheng_large_2020,baydin_magnetic_2022,hernandez_observation_2023}. Phenomenological extensions to Fermi-surface electrons have been proposed in Dirac semimetals~\cite{chen_gauge_2025,chen_geometric_2025}, but their model dependence limits their applicability. Together, these results indicate that essential physical ingredients are missing in the adiabatic description, particularly in the presence of finite carrier density or when phonon frequencies are comparable to electronic energy scales.

    In this Letter, we resolve this issue by developing a nonadiabatic theory of phonon magnetic moments that treats insulators and metals on equal footing. Our approach expresses the phonon magnetic moment as a linear response of ionic forces to velocities in a magnetic field and derives a gauge-invariant formula using a gauge-covariant Wigner expansion. The resulting expression naturally separates Fermi-sea and Fermi-surface contributions and retains the full frequency dependence of phonons. In the low-frequency limit for gapped systems, it reduces to previous adiabatic theories~\cite{trifunovic_geometric_2019,ren_phonon_2021,zhang_gate-tunable_2023}. 
    Beyond this limit, it captures additional contributions arising from nonadiabatic effects and from the Fermi surface.
    
    To illustrate these effects, we apply our theory to a low-energy model of the topological crystalline insulator Pb$_{1-x}$Sn$_x$Te. We find that nonadiabatic corrections are substantial even in the insulating regime when the electronic gap is narrow. Upon doping into the metallic regime, Fermi-surface electrons contribute significantly and eventually dominate the phonon magnetic moment. Using model parameters consistent with experimental measurements~\cite{hernandez_observation_2023}, we show that the Fermi-surface contribution plays a central role, leading to a phonon magnetic moment reaching the order of $\mu_B$, comparable to experimental observations. Our framework thus provides a systematic route for studying phonon magnetic moments in general systems, including both insulators and metals.

	\textit{Chiral phonon splitting.}---Conventionally, ionic motion is described under the assumption that time-reversal symmetry (TRS) is preserved.
	Within the harmonic approximation, it is governed by the dynamical matrix $D(\bm{q})$, whose eigenvalue $\omega_{\bm{q}\lambda}^2$ is the squared frequency of the $\lambda$-th phonon branch with quasi-wavevector $\bm{q}$. 
	The corresponding eigenmode $\bm{e}_{\bm{q}\lambda}$ describes the vibrating pattern $\bm{u}_{\lambda}(\bm{q},t) \propto \bm{e}_{\bm{q}\lambda} \,\mathrm{exp}(\mathrm{i}\bm{q}\cdot\bm{R}-\mathrm{i}\omega_{\bm{q}\lambda} t)$ of ions~\cite{giustino_electron-phonon_2017}, where $\bm{R}$ denotes the equilibrium positions of ions and $\bm{u}$ is the displacement.
	
	This description is modified when TRS is broken.  An additional Lorentz-like force emerges, which couples ionic displacements with their velocities~\cite{qin_berry_2012,zabalo_rotational_2022}. 
	The equation of motion then becomes $\dot{\hat{\bm{P}}}(\bm{q})=-\hat{D}(\bm{q}) \hat{\bm{u}}(\bm{q}) + \hat{G}(\bm{q},\omega) \hat{\bm{P}}(\bm{q})$, where $\hat{\bm P}(\bm q)$ is the ion momentum, and $\hat{G}(\bm{q},\omega)$ is a frequency-dependent anti-Hermitian matrix representing the Lorentz-like force. 
	This term lifts the degeneracy of phonon modes and gives rise to chiral phonons.
	To see this, consider two degenerate linearly polarized modes $\bm{e}_x(\bm{q})$ and $\bm{e}_y(\bm{q})$. In this subspace, $\hat{D}(\bm{q})=\omega_0^2(\bm{q})\,\hat{I}$ and
	\begin{eqnarray}
		G(\bm{q},\omega)
		=
		\begin{bmatrix}
			0 & g(\bm{q},\omega) \\
			-g^{*}(\bm{q},\omega) & 0
		\end{bmatrix}.
	\end{eqnarray}
	Solving the equation of motion, the eigenmodes become
 $\bm{e}_{R(L)}(\bm{q})=\left(\bm{e}_{x}(\bm{q})\pm \mathrm{i}\, \bm{e}_{y}(\bm{q})\right)/\sqrt{2} + O(g(\bm{q}))$,
	corresponding to right- and left-handed circularly polarized vibrations.  In the weak-coupling limit $|g(\bm{q},\omega_0)| \ll \omega_0(\bm{q})$, the resulting chiral-phonon splitting is
	\begin{eqnarray}
		\Delta \omega = | g(\bm{q},\omega_0(\bm{q})) |.
	\end{eqnarray}
    Under a magnetic field $\bm{B}$, the term proportional to $\bm{B}$ directly defines the phonon magnetic moment.
    In the following, we will consider general cases that the matrices $\hat{G}$ and $\hat{D}$ contain all ionic degrees of freedom, and focus on the TRS-breaking effect in the matrix $\hat G(\bm q, \omega)$. 

\begin{figure}
		\includegraphics[width=8.6cm]{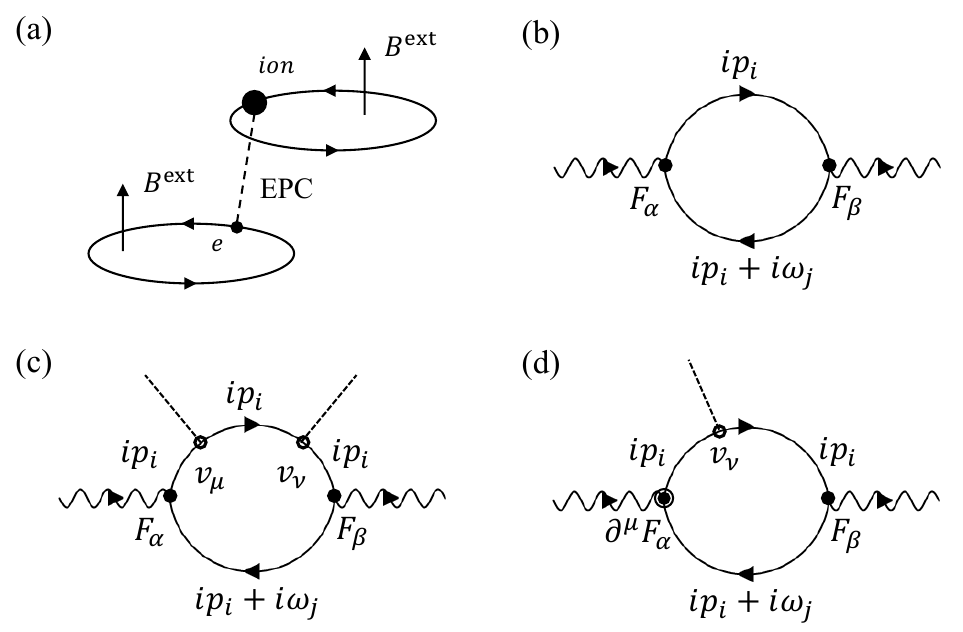}
		\caption{\label{fig:Feynman} (a) Schematic demonstration of how phonons couple to external magnetic fields, i.e., direct Lorentz-coupling of ions to $\bm{B}_{\mathrm{ext}}$ and coupling through EPC to electrons under magnetic fields.
			(b) Feynman diagrams for calculating chiral phonon splitting. (c) and (d) Feynman diagrams for calculating phonon magnetic moments.}
	\end{figure}

    In lattice dynamics, the TRS-breaking effects come from two sources (Fig.~\ref{fig:Feynman}\,(a)). The first is the direct coupling of ionic charges to an external magnetic field, which produces a splitting via the Lorentz force~\cite{juraschek_dynamical_2017,sun_phonon_2021}
	\begin{eqnarray}
		{G}^{(L)}_{\kappa a,\kappa'b}(\bm{q})= \sum_c \delta_{\kappa\kappa'} \frac{Z_\kappa e}{2M_\kappa}
		\epsilon_{abc} B_c ,
	\end{eqnarray}
	where $Z_\kappa e\equiv -Z_\kappa |e|$ is the bare charge of ion $\kappa$, $M_\kappa$ is its mass and $\bm{B}$ is the magnetic field. The Levi-Civita tensor $\epsilon_{abc}$ denotes antisymmetric permutations of Cartesian directions $a$, $b$ and $c$.  

    The second, and equally important, source arises from electron-phonon coupling (EPC), where time-reversal symmetry (TRS) is broken in the electronic sector, either by an external magnetic field or intrinsic magnetic order. A key step is to formulate this contribution in a linear-response framework~\cite{liu_circular_2017,wang_ab_2025}, by identifying $\hat{G}$ as the response function relating the force  $\hat{\bm{F}}_{\kappa}(\bm{q})=-\sum_l e^{\mathrm{i}\bm{q}\cdot\bm{R}_l} {\partial \hat{H}}/{\partial \bm{R}_{l\kappa}}$ acting on ion $\kappa$ to the velocity $\dot{\bm{u}}_{\kappa'}(\bm{q})$ of another ion $\kappa'$.
	Within this formulation, $\hat{G}$ naturally takes the form of a Kubo-type response function. By noticing that $\dot{\bm{u}}_{\kappa'}= -\mathrm{i}\omega\bm{u}_{\kappa'}$, for a non-interacting system, we obtain~\cite{giuliani_quantum_2005}
    \begin{multline}\label{G_EPC}
		G^{(\mathrm{EPC})}_{\kappa a,\kappa'b}(\bm{q},\omega)
		= \frac{-\mathrm{i}\hbar}{2\omega\sqrt{M_\kappa M_{\kappa'}}} 
		\Bigg\{
		\frac{1}{\hbar\beta}\sum_i
		\mathrm{Tr}\,
		\bm{\Big[}
		\hat{F}_{\kappa a}(\bm{q})\,
		\\
		\hat{\mathcal{G}}(\bm{k},\mathrm{i}p_i)\,
		\hat{F}_{\kappa'b}(-\bm{q})\,
		\hat{\mathcal{G}}(\bm{k+q},\mathrm{i}p_i+\mathrm{i}\omega_j)\,
		\bm{\Big]}
		\Bigg\}_{\substack{\mathrm{i}\omega_j \\ \rightarrow \omega+\mathrm{i}\eta}}\\
		-(\kappa a \leftrightarrow \kappa' b).\quad\quad\quad
	\end{multline}
	Here $\hat{\mathcal{G}}(\bm{k},\mathrm{i}p_i)$ stands for electron Green's functions with quasi-wavevector $\bm{k}$ and fermionic Matsubara frequency $\mathrm{i}p_i=\mathrm{i}(2i+1)\pi/\hbar\beta$, with $\beta=1/k_B T$ being the inverse temperature. 
	The wavevector $\bm{q}$ and frequency transfer $\mathrm{i}\omega_j$ correspond to those of phonons absorbed or emitted in the process.
    The notation $\mathrm{i}\omega_j \rightarrow \omega+\mathrm{i}\eta$ denotes an analytical continuation after summation over $p_i$, which gives the dependence of $\hat{G}$ on phonon frequency $\omega$.
	The corresponding Feynman diagram is shown in Fig.~\ref{fig:Feynman}\,(b).
    
    Formulating $\hat{G}^{(\mathrm{EPC})}$ as a linear response function provides a unified framework on chiral phonon splitting and phonon magnetic moments.
    First, it provides a systematic route for incorporating additional perturbations and many-body effects into $G^{(\mathrm{EPC})}$.
    In particular, the phonon magnetic moment studied here can be obtained by introducing magnetic-field perturbations. 
    It also allows disorder and Coulomb-interaction effects to be included by replacing the single-particle linear response function with the disorder-averaged or interaction-dressed one.
    Second, it naturally builds a bridge between several earlier theories. When TRS breaking is intrinsic, Eq.~\eqref{G_EPC} reduces to previously developed nonadiabatic theories in magnetic systems~\cite{liu_circular_2017,bistoni_intrinsic_2021,wang_ab_2025}. In the adiabatic limit $\omega\rightarrow 0$, it further reduces to the molecular Berry curvature obtained within the Born-Oppenheimer approximation~\cite{mead_geometric_1992,qin_berry_2012,saparov_lattice_2022,bonini_frequency_2023}, corresponding to the geometric phase accumulated by electrons under cyclic ionic motion.  
    Furthermore, in the presence of magnetic fields,
    Eq.~\eqref{G_EPC} naturally connects to the recently developed gauge theory of phonon magnetic moments in Dirac semimetals~\cite{chen_gauge_2025}, where ${\partial \hat{H}}/{\partial \bm{R}} \propto {\partial \hat{H}}/{\partial \bm{k}}$. In this case, the force operator becomes proportional to the electronic velocity operator, implying that the chiral phonon splitting under magnetic fields scales with the Hall conductivity $\sigma_{xy}$. In the following, we make no assumptions about the form of the force operator and focus on nonmagnetic materials, where TRS is broken by an external magnetic field.

    \textit{Phonon magnetic moment.}---To evaluate the phonon magnetic moment, we calculate the phonon splitting to linear order in $\bm{B}$.
    More specifically, we need to calculate 
    $\gamma^{\mathrm{(EPC)}}_{\kappa a,\kappa'b,c}\equiv\hbar \partial G^{\mathrm{(EPC)}}_{\kappa a,\kappa'b}/\partial B_c$, where $c$ labels the cartesian component of $\bm{B}$.
    For the two-mode model introduced earlier, this linear coefficient directly determines the phonon magnetic moment. 
    For more general systems, one substitutes $G^{(\mathrm{EPC})} = \bm{\gamma}^{(\mathrm{EPC})}\cdot\bm{B}$ back into the equation of motion and solves for the resulting phonon eigenfrequencies, from which the splitting, and hence the magnetic moment, can be obtained.
    
    To include the effect of magnetic fields in a gauge-invariant way, we transform Eq.~(\ref{G_EPC}) to the Wigner representation~\cite{onoda_theory_2006,zhu_theory_2012,shitade_gradient_2017,noauthor_see_nodate}. In this representation, the magnetic field modifies operator products according to $\hat{A}_1\hat{A}_2 \Rightarrow \hat{A}_1\hat{A}_2 + \sum_c\frac{\mathrm{i} eB_c\hbar}{2}\epsilon_{\mu\nu c} (\partial_{p_\mu}\hat{A}_1) (\partial_{p_\nu}\hat{A}_2)$
	up to the linear order $\bm B$, where $\mu,\nu=x,y,z$ denotes the Cartesian direction of momentum $p$.
	This leads to two types of corrections to $G^{(\mathrm{EPC})}$.  
	The first type is to the Green's function, leading to $\hat{\mathcal{G}}^B
	\approx
	\hat{\mathcal{G}}
	+
	\sum_c
	\frac{i eB_c\hbar }{2} \epsilon_{\mu\nu c}\,
	\hat{\mathcal{G}}\,
	\hat{v}_\mu\,
	\hat{\mathcal{G}}\,
	\hat{v}_\nu\,
	\hat{\mathcal{G}}$
	for a non-interacting system~\cite{onoda_theory_2006,shitade_gradient_2017}. 
	The second type of corrections is to the product between Green's functions and force operators by replacing $\hat{A}_1$ and $\hat{A}_2$ above with $\hat{\mathcal{G}}$ and $\hat{F}$. Details can be found in the Supplemental Materials (SM)~\cite{noauthor_see_nodate}.
	Combining these contributions, we obtain
	\begin{multline}\label{gamma_EPC}
		\gamma^{(\mathrm{EPC})}_{\kappa a,\kappa' b,c}(\omega)
		=\frac{e \hbar^2 \epsilon_{\mu\nu c}}{4\omega\sqrt{M_\kappa M_{\kappa'}}}
		\Bigg\{\frac{1}{\hbar\beta}\sum_{i} \bm{\big[}\kappa a,\mu,\nu,\kappa'b\bm{\big]}_{i+j}\\
		+ \bm{\big[}(\kappa a,\mu),\nu,\kappa'b\bm{\big]}_{i+j} - (j\rightarrow -j)\Bigg\}_{\mathrm{i}\omega_j\rightarrow\omega+\mathrm{i}\eta}\\
		-(\kappa a \leftrightarrow \kappa' b),\quad\quad\quad\quad
	\end{multline}
	where we define $\big[\kappa a,\mu,\nu,\kappa'b\big]_{i+j}\equiv$$\mathrm{Tr}
	\Big[\hat{F}_{\kappa a}\,
	\hat{\mathcal{G}}(\mathrm{i}p_i)\,
	\hat{v}_\mu\,
	\hat{\mathcal{G}}(\mathrm{i}p_i)\,$
	$
	\hat{v}_\nu\,
	\hat{\mathcal{G}}(\mathrm{i}p_i)\,
	\hat{F}_{\kappa' b}\,
	\hat{\mathcal{G}}(\mathrm{i}p_i+\mathrm{i}\omega_j)\Big]$
	and
	$\big[(\kappa a,\mu),\nu,\kappa'b\big]_{i+j}\equiv
	\mathrm{Tr}\Big[
	(\partial^{\kappa a}\hat{v}_\mu)\,
	\hat{\mathcal{G}}(\mathrm{i}p_i)\,
	\hat{v}_\nu\,
	\hat{\mathcal{G}}(\mathrm{i}p_i)\,
	\hat{F}_{\kappa' b}\,
	\hat{\mathcal{G}}(\mathrm{i}p_i+\mathrm{i}\omega_j)\Big]$.
    The corresponding Feynman diagrams are shown in Figs.~\ref{fig:Feynman}\,(c) and\,(d).
	The second term exists only when the non-kinetic part of the Hamiltonian is non-local, so that $\partial^{\kappa\alpha} \hat{v}_\mu \equiv \partial \hat{v}_\mu/\partial \bm{R}_{\kappa\alpha}\ne0$. In DFT, this is generally the case because pseudopotentials are non-local, so we need to keep both terms for completeness. 
    
	After completing the Matsubara frequency summation and analytical continuation, we obtain the formula for phonon magnetic moments in general systems. 
	The formula contains not only Fermi-sea terms, but also Fermi-surface contributions, i.e.,
	\begin{eqnarray}
        {\gamma}^{(\mathrm{EPC})}_{\kappa a,\kappa' b,c} = {\gamma}^{(\mathrm{EPC,sea})}_{\kappa a,\kappa' b,c} + {\gamma}^{(\mathrm{EPC,FS})}_{\kappa a,\kappa' b,c}.
	\end{eqnarray}
	As a result, it can be applied to both metals and insulators.
	In an insulator, the Fermi-surface term vanishes. 
	We find that the frequency dependence of the formula comes entirely from the inter-gap transition process~\cite{noauthor_see_nodate}.
	In the low-frequency insulating limit ($\hbar\omega\ll\Delta_{\mathrm{gap}}$),  
    the formula reduces exactly to the adiabatic results of Refs.~\cite{ren_phonon_2021,zhang_gate-tunable_2023,trifunovic_geometric_2019}. 
	The complete formula and proof can be found in the SM~\cite{noauthor_see_nodate}.
	The current theory, however, does not rely on the adiabatic approximation and captures the full frequency-dependence of the phonon magnetic moment, making the theory applicable even when $\hbar\omega > \Delta_{\mathrm{gap}}$.
	
	The Fermi-surface contribution, absent in previous microscopic theories, is given by
	\begin{widetext}
		\begin{eqnarray}\label{phmag_FS}
			\gamma^{(\mathrm{EPC,FS})}_{\kappa a,\kappa'b,c}(\omega)
			&&=\frac{e \hbar^2 \epsilon_{\mu\nu c}}{4\omega\sqrt{M_\kappa M_{\kappa'}}}
			\sum_{nmlp}
			(F_{\kappa a})_{nm} (v_\mu)_{ml} (v_\nu)_{lp} (F_{\kappa'b})_{pn}\nonumber\\
			&&\times\Bigg\{\frac{f_m'}{\Delta_{mp}(\Delta_{mn}+\hbar\omega+\mathrm{i}\Gamma)} \delta_{\varepsilon_m,\varepsilon_l}\bar{\delta}_{\varepsilon_l,\varepsilon_p}
			+ (\mathrm{cyclic\ of\ } m,l,p)
			- (\omega + \mathrm{i}\Gamma \rightarrow -\omega - \mathrm{i}\Gamma)
			\Bigg\}\nonumber\\
			&&+
			\frac{e\hbar\epsilon_{\mu\nu c}}{4\omega\sqrt{M_\kappa M_{\kappa'}}}
			\sum_{nmlp}
			(\partial^{\kappa a} v_\mu)_{nm} (v_\nu)_{ml} (F_{\kappa'b})_{ln}
			\Bigg\{\frac{f_m'}{\Delta_{mn}+\hbar\omega+\mathrm{i}\Gamma} \delta_{\varepsilon_m,\varepsilon_l}
			- (\omega + \mathrm{i}\Gamma \rightarrow -\omega - \mathrm{i}\Gamma)
			\Bigg\}\nonumber\\
			&&\nonumber\\
			&&-(\kappa a \leftrightarrow \kappa' b),
		\end{eqnarray}
	\end{widetext}
	where $\Delta_{mp}=\varepsilon_m - \varepsilon_{p}$ is the energy difference between the bands, $\delta_{\varepsilon_m,\varepsilon_l}=1$ when $\varepsilon_m=\varepsilon_l$ and otherwise $0$, $\bar{\delta}_{\varepsilon_m,\varepsilon_l}\equiv 1 -\delta_{\varepsilon_m,\varepsilon_l}$ and we use  $(\mathrm{cyclic\ of\ } m,l,p)$ to denote the two additional terms obtained by cyclic permutations $(m,l,p)\rightarrow(l,p,m)\rightarrow(p,m,l)$. 
	One may notice that Fermi-surface contribution for $\Delta_{mn}=0$ diverges when $\omega\rightarrow0$. This is regulated by introducing a finite electron lifetime by replacing the infinitesimal $\eta$ with a finite $\Gamma$ (see SM for details~\cite{noauthor_see_nodate}).
	In practice, the linewidth $\Gamma$ can be induced by EPC, disorder and so on. 
	To this end, we have completed the construction of a nonadiabatic theory of phonon magnetic moment applicable to both insulators and metals.


	\begin{figure}[t!]
		\includegraphics[width=8.6cm]{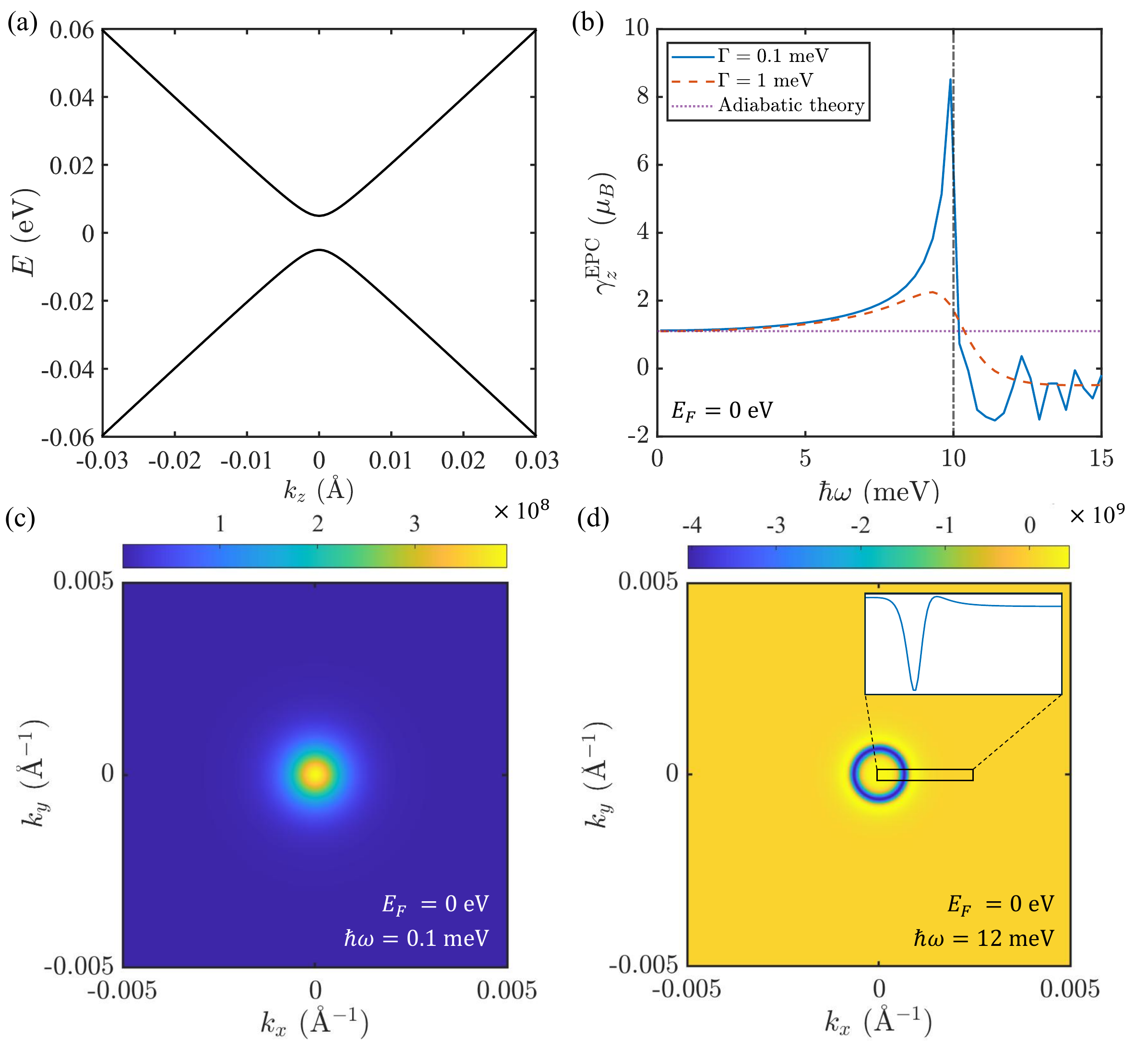}
		\caption{\label{fig:phmag}
			(a) Electronic dispersion along $k_3$ at $k_1, k_2=0$. The gap size is $\Delta_{\mathrm{gap}}=2m=10$ meV. Other parameters are set as $c=0.9$ eV, $f=4.5$ eV, $v_3=0.95$ eV, $v_F=2.4$ eV~\cite{lau_topological_2019,hernandez_observation_2023}, $\zeta=0.5$ eV/\AA, $\Gamma=1$ meV. The phonon mass is set to the reduced mass $M_{\mathrm{Pb}}M_{\mathrm{Te}}/(M_{\mathrm{Pb}}+M_{\mathrm{Te}})\approx 79$ Da.
			(b) Dependence of phonon magnetic moment on the phonon frequency $\hbar\omega$. 
            Results using both Eq.~(\ref{gamma_EPC}) in this work and adiabatic theory in Ref.~\cite{zhang_gate-tunable_2023,ren_phonon_2021,trifunovic_geometric_2019} are shown for comparison. 
            The vertical dashed line denotes the energy where $\hbar\omega=\Delta_{\mathrm{gap}}$.
			(c-d) Distribution of phonon magnetic moment in the reciprocal space at $k_z=0$. 
			Phonon frequencies are set to the adiabatic limit ($\hbar\omega=0.1$ meV) and values exceeding the gap size ($\hbar\omega=12$ meV), respectively. The inset in (e) shows the distribution along $k_x$ at $k_y=0$.
		}
	\end{figure}
	
	Importantly, the formulas above suggest two processes that could generate phonon magnetic moments substantially larger than the conventional adiabatic term in insulators.
	The first is the inter-gap resonant process mediated by phonons in the Fermi-sea contribution. This involves virtual transitions between occupied and unoccupied states whose energy difference is close to $\hbar\omega$.
	The corresponding terms take the form of $(\Delta_{nm} + \hbar\omega +\mathrm{i}\Gamma)^{-r}$, and (near) resonance leads to enhancements proportional to the inverse of the electronic linewidth $1/(\Gamma)^{r}$, where $r=1$ or $2$.
	The second mechanism is the intraband process on the Fermi surface, which contributes terms proportional to $1/(\hbar\omega+i\Gamma)^2$ in the small-$\omega$ limit. \textcolor{black}{In the absence of interband resonance, such as in a single-band system (e.g., free electrons) or doped Dirac semimetals where $2E_F \gg \hbar\omega$, this intraband Fermi-surface contribution dominates the phonon magnetic moment.}
	In addition, it is worth noting that the behavior is very similar to that of optical Hall conductivity obtained from Drude theory~\cite{hartnoll_holographic_2018}.

    \textit{Application to Doped Pb$_{1-x}$Sn$_{x}$Te.}---We now apply our theory to a massive anisotropic Dirac $k\cdot p$ model described by the Hamiltonian $\hat{H}=\hat{H}_0+\hat{H}_{\mathrm{EPC}}$, where $\hat{H}_0=(m+ck_3^2+fk_1^2+fk_2^2)\hat{s}_0\hat{\sigma}_z+v_F(k_1 \hat{s}_y-k_2 \hat{s}_x)\hat{\sigma}_x + v_3 k_3 \hat{s}_0 \hat{\sigma}_y$ with the sign of the gap $2m$ controlling the topological order, and the coupling to phonons is given by $\hat{H}_{\mathrm{EPC}}=\zeta (u_x \hat{s}_0 \hat{\sigma}_x - u_y \hat{s}_z \hat{\sigma}_y)$. Both $\hat{s}$ and $\hat{\sigma}$ are Pauli matrices, denoting spin and orbital degrees of freedom, respectively. This model can be used to describe Pb$_{1-x}$Sn$_{x}$Te~\cite{lau_topological_2019,hernandez_observation_2023}. 
    In the calculation below, the electron linewidth $\Gamma$ is set to $1$ meV unless stated explicitly. 
	Using this model, we demonstrate how finite phonon frequency and finite carrier density can amplify and even reverse the EPC-induced phonon magnetic moment. 
	
	We begin with the narrow-gap-insulator case to demonstrate the nonadiabatic effect. The gap is set to be $\Delta_{\mathrm{gap}}=2m=10$ meV and the Fermi level is placed within the gap. 
    In Fig.~\ref{fig:phmag}(b), the dotted horizontal line shows the frequency-independent adiabatic formula in Ref.~\cite{trifunovic_geometric_2019,ren_phonon_2021,zhang_gate-tunable_2023}.  We can see that when $\hbar\omega \ll \Delta$, our result agrees well with the adiabatic theory.  When $\hbar\omega$ approaches the gap size, the phonon magnetic moment is strongly enhanced because of the near-resonance between inter-gap states.  
    When $\hbar\omega$ exceeds the gap size, it undergoes an abrupt change and even reverses sign as $\hbar\omega$ further increases.
	The results indicate the important role of nonadiabatic effects in determining phonon magnetic moment.
	
	The pronounced frequency dependence originates from these near-resonant processes. In Figs.~\ref{fig:phmag}(c,d), we show the momentum-resolved distribution of the phonon magnetic moment. For $\hbar\omega < \Delta_{\mathrm{gap}}$ [Fig.~\ref{fig:phmag}(c)], no exact resonance occurs, but contributions become increasingly concentrated near $\bm{k}=0$ as the phonon frequency increases and resonance is approached. For example, at $\hbar\omega=8$ meV, the distribution remains centered near $\bm{k}=0$ while its magnitude increases by a factor of four (Fig.~S1). When $\hbar\omega > \Delta_{\mathrm{gap}}$ (Fig.~\ref{fig:phmag}(d)), resonance sets in, producing sharp features in momentum space and accompanying a sign reversal of the phonon magnetic moment.

	\begin{figure}[t!]
		\includegraphics[width=8.6cm]{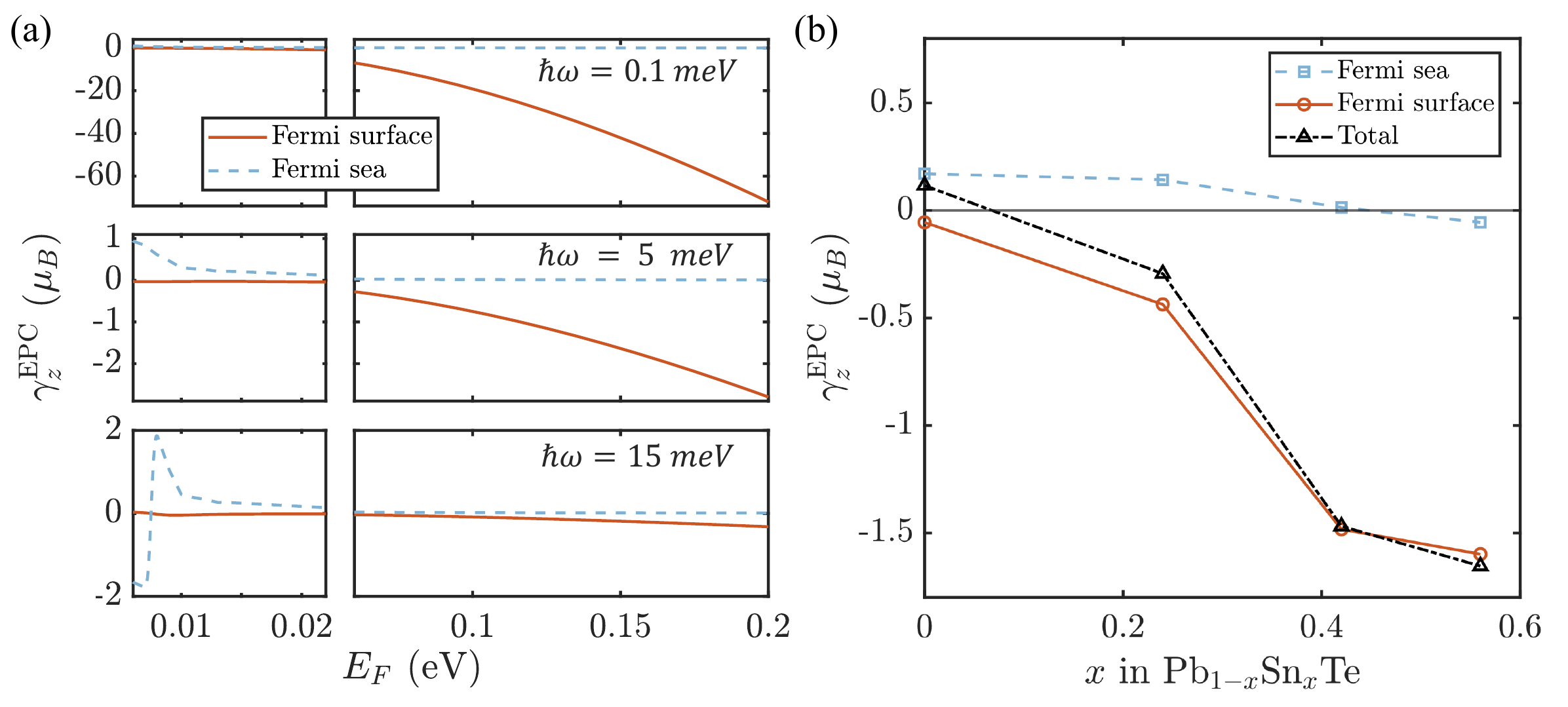}
		\caption{\label{fig:phmag_FS}
			(a) Dependence of Fermi-surface (solid lines) and Fermi-sea (dashed lines) contributions to the phonon magnetic moment on the position of the Fermi level $E_F$. Results for phonon at $\hbar\omega=0.1$ meV (top), $5$ meV (middle) and $15$ meV (bottom) are shown for comparison. The $E_F$-axis is split into low-doping and high-doping level region for clarity. The gap size is $\Delta_{\mathrm{gap}}=2m=10$ meV.
			(b) Dependence of the magnetic moment of a $5$-meV phonon on the Sn concentration $x$ in Pb$_{1-x}$Sn$_{x}$Te. Gap sizes and positions of the Fermi level are determined according to the hole concentrations measured in experiments~\cite{hernandez_observation_2023}. Parameters are set to $\zeta=0.5$ eV/\AA\ and $\Gamma=10$ meV. The phonon mass is set to the reduced mass $M_{\mathrm{Pb}_{1-x}\mathrm{Sn}_{x}}M_{\mathrm{Te}}/(M_{\mathrm{Pb}_{1-x}\mathrm{Sn}_{x}}+M_{\mathrm{Te}})$.
		}
	\end{figure}
	
	When the system is doped into a metallic regime, an additional Fermi-surface contribution emerges. Figure~\ref{fig:phmag_FS}(a) shows both Fermi-sea and Fermi-surface contributions to the phonon magnetic moment as functions of the Fermi level $E_F$ at different phonon frequencies. At low doping, the response is dominated by the Fermi-sea term due to the small number of Fermi-surface electrons. As $E_F$ increases, the growing density of states at the Fermi level enhances the Fermi-surface contribution, which rapidly becomes dominant.  This Fermi-surface term exhibits a strong frequency dependence, scaling approximately as $(\hbar\omega+i\Gamma)^{-2}$ as discussed above. Consequently, in the low-phonon-frequency limit, it can exceed the Fermi-sea contribution by up to two orders of magnitude. At higher phonon frequencies (e.g., $\hbar\omega=15$ meV), interband resonance becomes active, leading to enhanced Fermi-sea contributions and a sign change in both components when $2E_F$ crosses $\hbar\omega$. These results highlight the central role of Fermi-surface contributions in determining phonon magnetic moments in metals.
    
    Next, we connect our results to recent experiments in Pb$_{1-x}$Sn$_x$Te, where the phonon magnetic moment was measured as a function of substitution level $x$. For each $x$, we set the Fermi level according to the experimentally measured hole concentration and use the corresponding gap size $m$. The magnetic moment is evaluated for phonons at $5$ meV~\cite{hernandez_observation_2023}.  Figure~\ref{fig:phmag_FS}(b) shows the calculated Fermi-sea and Fermi-surface contributions as functions of $x$. The total $\gamma^{\mathrm{EPC}}$ increases markedly with Sn concentration. Using an EPC strength $\zeta \sim 0.5$ eV/\AA\ fitted from DFT~\cite{noauthor_see_nodate}, a scattering rate $\Gamma \sim 10$ meV estimated from mobility measurements~\cite{assaf_magnetooptical_2017}, and including contributions from all four valleys, we obtain $|\gamma^{\mathrm{(EPC)}}| \approx 0.11\,\mu_B$ ($x=0$), $0.29\,\mu_B$ ($x=0.24$), and $1.47\,\mu_B$ ($x=0.42$). These values are in reasonable agreement with the experimental measurements ($0.04\,\mu_B$, $0.12\,\mu_B$, and $1\,\mu_B$, respectively, and both experimentally resolved branches (see below) have comparable magnitudes~\cite{hernandez_observation_2023}). At low doping, the Fermi-sea and Fermi-surface contributions are comparable, but the latter quickly becomes dominant as $x$ increases.

    Despite this quantitative agreement, some experimental features are not captured. In particular, the two experimentally resolved phonon branches behave differently across the topological crystalline insulator transition near $x_c \approx 0.32$: one exhibits a sign change, while the other does not. This branch-dependent behavior is absent in our calculation due to the harmonic approximation. Since the primitive cell contains two atoms, the harmonic theory supports only a single transverse optical (TO) doublet; the two observed branches arise from anharmonic splitting of this mode~\cite{delaire_giant_2011,ribeiro_strong_2018,hernandez_observation_2023}. Capturing this effect requires going beyond the harmonic approximation.
	
	
	\textit{Summary.}---We have developed a nonadiabatic framework for phonon magnetic moments that applies to both insulating and metallic systems. The theory identifies two key contributions beyond the adiabatic limit: resonant interband processes and an intraband Fermi-surface term that can become dominant in metals, leading to substantial corrections to previous adiabatic estimates. As a result, the phonon magnetic moment acquires a strong dependence on both frequency and carrier density. Application to Pb$_{1-x}$Sn$_x$Te shows that these effects account for the experimentally observed magnitude and trends, placing phonon magnetic moments on the order of $\mu_B$. Extending the present framework to incorporate anharmonic phonon effects remains an important direction for future work.
	
	\textit{Acknowledgment.}---We gratefully acknowledge Xiaodong Hu, Xiaoyu Liu, Junren Shi, Zezhu Wei, and Xiao-Wei Zhang for valuable discussions. 
    The theory development is supported by DOE Award No.~DE-SC0012509, and the first-principles calculations and materials modeling are supported by NSF Award No. DMR-2339995. 
    
    \nocite{ismail-beigi_coupling_2001,chen_unified_2011,giannozzi_quantum_2009,giannozzi_advanced_2017,hamann_optimized_2013,perdew_generalized_1996}

	\bibliographystyle{apsrev4-2}
	\bibliography{Reference}

@article{qin_berry_2012,
	title = {Berry curvature and the phonon {Hall} effect},
	volume = {86},
	url = {https://link.aps.org/doi/10.1103/PhysRevB.86.104305},
	doi = {10.1103/PhysRevB.86.104305},
	number = {10},
	urldate = {2021-01-25},
	journal = {Physical Review B},
	author = {Qin, Tao and Zhou, Jianhui and Shi, Junren},
	month = sep,
	year = {2012},
	note = {Publisher: American Physical Society},
	pages = {104305},
}

@article{hamann_optimized_2013,
	title = {Optimized norm-conserving {Vanderbilt} pseudopotentials},
	volume = {88},
	url = {https://link.aps.org/doi/10.1103/PhysRevB.88.085117},
	doi = {10.1103/PhysRevB.88.085117},
	number = {8},
	urldate = {2021-01-25},
	journal = {Physical Review B},
	author = {Hamann, D. R.},
	month = aug,
	year = {2013},
	note = {Publisher: American Physical Society},
	pages = {085117},
}

@article{giannozzi_advanced_2017,
	title = {Advanced capabilities for materials modelling with {Quantum} {ESPRESSO}},
	volume = {29},
	issn = {0953-8984},
	url = {https://doi.org/10.1088/1361-648x/aa8f79},
	doi = {10.1088/1361-648X/aa8f79},
	language = {en},
	number = {46},
	urldate = {2021-01-25},
	journal = {Journal of Physics: Condensed Matter},
	author = {Giannozzi, P. and Andreussi, O. and Brumme, T. and Bunau, O. and Nardelli, M. Buongiorno and Calandra, M. and Car, R. and Cavazzoni, C. and Ceresoli, D. and Cococcioni, M. and Colonna, N. and Carnimeo, I. and Corso, A. Dal and Gironcoli, S. de and Delugas, P. and DiStasio, R. A. and Ferretti, A. and Floris, A. and Fratesi, G. and Fugallo, G. and Gebauer, R. and Gerstmann, U. and Giustino, F. and Gorni, T. and Jia, J. and Kawamura, M. and Ko, H.-Y. and Kokalj, A. and Küçükbenli, E. and Lazzeri, M. and Marsili, M. and Marzari, N. and Mauri, F. and Nguyen, N. L. and Nguyen, H.-V. and Otero-de-la-Roza, A. and Paulatto, L. and Poncé, S. and Rocca, D. and Sabatini, R. and Santra, B. and Schlipf, M. and Seitsonen, A. P. and Smogunov, A. and Timrov, I. and Thonhauser, T. and Umari, P. and Vast, N. and Wu, X. and Baroni, S.},
	month = oct,
	year = {2017},
	note = {Publisher: IOP Publishing},
	pages = {465901},
}

@article{giustino_electron-phonon_2017,
	title = {Electron-phonon interactions from first principles},
	volume = {89},
	url = {https://link.aps.org/doi/10.1103/RevModPhys.89.015003},
	doi = {10.1103/RevModPhys.89.015003},
	number = {1},
	urldate = {2021-01-25},
	journal = {Reviews of Modern Physics},
	author = {Giustino, Feliciano},
	month = feb,
	year = {2017},
	note = {Publisher: American Physical Society},
	pages = {015003},
}

@article{chen_enhanced_2020,
	title = {Enhanced {Thermal} {Hall} {Effect} in {Nearly} {Ferroelectric} {Insulators}},
	volume = {124},
	url = {https://link.aps.org/doi/10.1103/PhysRevLett.124.167601},
	doi = {10.1103/PhysRevLett.124.167601},
	number = {16},
	urldate = {2021-05-29},
	journal = {Physical Review Letters},
	author = {Chen, Jing-Yuan and Kivelson, Steven A. and Sun, Xiao-Qi},
	month = apr,
	year = {2020},
	note = {Publisher: American Physical Society},
	pages = {167601},
}

@article{li_phonon_2020,
	title = {Phonon {Thermal} {Hall} {Effect} in {Strontium} {Titanate}},
	volume = {124},
	url = {https://link.aps.org/doi/10.1103/PhysRevLett.124.105901},
	doi = {10.1103/PhysRevLett.124.105901},
	number = {10},
	urldate = {2021-05-30},
	journal = {Physical Review Letters},
	author = {Li, Xiaokang and Fauqué, Benoît and Zhu, Zengwei and Behnia, Kamran},
	month = mar,
	year = {2020},
	note = {Publisher: American Physical Society},
	pages = {105901},
}

@article{boulanger_thermal_2020,
	title = {Thermal {Hall} conductivity in the cuprate {Mott} insulators {Nd2CuO4} and {Sr2CuO2Cl2}},
	volume = {11},
	issn = {2041-1723},
	url = {http://www.nature.com/articles/s41467-020-18881-z},
	doi = {10.1038/s41467-020-18881-z},
	language = {en},
	number = {1},
	urldate = {2021-05-31},
	journal = {Nature Communications},
	author = {Boulanger, Marie-Eve and Grissonnanche, Gaël and Badoux, Sven and Allaire, Andréanne and Lefrançois, Étienne and Legros, Anaëlle and Gourgout, Adrien and Dion, Maxime and Wang, C. H. and Chen, X. H. and Liang, R. and Hardy, W. N. and Bonn, D. A. and Taillefer, Louis},
	month = dec,
	year = {2020},
	pages = {5325},
}

@article{zhang_thermal_2019,
	title = {Thermal {Hall} {Effect} {Induced} by {Magnon}-{Phonon} {Interactions}},
	volume = {123},
	url = {https://link.aps.org/doi/10.1103/PhysRevLett.123.167202},
	doi = {10.1103/PhysRevLett.123.167202},
	number = {16},
	urldate = {2021-06-12},
	journal = {Physical Review Letters},
	author = {Zhang, Xiaoou and Zhang, Yinhan and Okamoto, Satoshi and Xiao, Di},
	month = oct,
	year = {2019},
	note = {Publisher: American Physical Society},
	pages = {167202},
}

@article{saito_berry_2019,
	title = {Berry {Phase} of {Phonons} and {Thermal} {Hall} {Effect} in {Nonmagnetic} {Insulators}},
	volume = {123},
	url = {https://link.aps.org/doi/10.1103/PhysRevLett.123.255901},
	doi = {10.1103/PhysRevLett.123.255901},
	number = {25},
	urldate = {2021-07-11},
	journal = {Physical Review Letters},
	author = {Saito, Takuma and Misaki, Kou and Ishizuka, Hiroaki and Nagaosa, Naoto},
	month = dec,
	year = {2019},
	note = {Publisher: American Physical Society},
	pages = {255901},
}

@article{sun_phonon_2021,
	title = {Phonon {Hall} effect with first-principles calculations},
	volume = {103},
	url = {https://link.aps.org/doi/10.1103/PhysRevB.103.214301},
	doi = {10.1103/PhysRevB.103.214301},
	number = {21},
	urldate = {2021-07-11},
	journal = {Physical Review B},
	author = {Sun, Kangtai and Gao, Zhibin and Wang, Jian-Sheng},
	month = jun,
	year = {2021},
	note = {Publisher: American Physical Society},
	pages = {214301},
}

@article{mead_geometric_1992,
	title = {The geometric phase in molecular systems},
	volume = {64},
	url = {https://link.aps.org/doi/10.1103/RevModPhys.64.51},
	doi = {10.1103/RevModPhys.64.51},
	number = {1},
	urldate = {2021-07-11},
	journal = {Reviews of Modern Physics},
	author = {Mead, C. Alden},
	month = jan,
	year = {1992},
	note = {Publisher: American Physical Society},
	pages = {51--85},
}

@article{ideue_giant_2017,
	title = {Giant thermal {Hall} effect in multiferroics},
	volume = {16},
	issn = {1476-1122, 1476-4660},
	url = {http://www.nature.com/articles/nmat4905},
	doi = {10.1038/nmat4905},
	language = {en},
	number = {8},
	urldate = {2021-08-29},
	journal = {Nature Materials},
	author = {Ideue, T. and Kurumaji, T. and Ishiwata, S. and Tokura, Y.},
	month = aug,
	year = {2017},
	pages = {797--802},
}

@article{perdew_generalized_1996,
	title = {Generalized {Gradient} {Approximation} {Made} {Simple}},
	volume = {77},
	url = {https://link.aps.org/doi/10.1103/PhysRevLett.77.3865},
	doi = {10.1103/PhysRevLett.77.3865},
	number = {18},
	urldate = {2021-11-19},
	journal = {Physical Review Letters},
	author = {Perdew, John P. and Burke, Kieron and Ernzerhof, Matthias},
	month = oct,
	year = {1996},
	note = {Publisher: American Physical Society},
	pages = {3865--3868},
}

@article{giannozzi_quantum_2009,
	title = {{QUANTUM} {ESPRESSO}: a modular and open-source software project for quantum simulations of materials},
	volume = {21},
	issn = {0953-8984},
	shorttitle = {{QUANTUM} {ESPRESSO}},
	url = {https://doi.org/10.1088/0953-8984/21/39/395502},
	doi = {10.1088/0953-8984/21/39/395502},
	language = {en},
	number = {39},
	urldate = {2021-11-19},
	journal = {Journal of Physics: Condensed Matter},
	author = {Giannozzi, Paolo and Baroni, Stefano and Bonini, Nicola and Calandra, Matteo and Car, Roberto and Cavazzoni, Carlo and Ceresoli, Davide and Chiarotti, Guido L. and Cococcioni, Matteo and Dabo, Ismaila and Corso, Andrea Dal and Gironcoli, Stefano de and Fabris, Stefano and Fratesi, Guido and Gebauer, Ralph and Gerstmann, Uwe and Gougoussis, Christos and Kokalj, Anton and Lazzeri, Michele and Martin-Samos, Layla and Marzari, Nicola and Mauri, Francesco and Mazzarello, Riccardo and Paolini, Stefano and Pasquarello, Alfredo and Paulatto, Lorenzo and Sbraccia, Carlo and Scandolo, Sandro and Sclauzero, Gabriele and Seitsonen, Ari P. and Smogunov, Alexander and Umari, Paolo and Wentzcovitch, Renata M.},
	month = sep,
	year = {2009},
	note = {Publisher: IOP Publishing},
	pages = {395502},
}

@article{bonini_frequency_2023,
	title = {Frequency {Splitting} of {Chiral} {Phonons} from {Broken} {Time}-{Reversal} {Symmetry} in {CrI3}},
	volume = {130},
	url = {https://link.aps.org/doi/10.1103/PhysRevLett.130.086701},
	doi = {10.1103/PhysRevLett.130.086701},
	number = {8},
	urldate = {2023-03-08},
	journal = {Physical Review Letters},
	author = {Bonini, John and Ren, Shang and Vanderbilt, David and Stengel, Massimiliano and Dreyer, Cyrus E. and Coh, Sinisa},
	month = feb,
	year = {2023},
	note = {Publisher: American Physical Society},
	pages = {086701},
}

@article{li_phonon_2023,
	title = {The phonon thermal {Hall} angle in black phosphorus},
	volume = {14},
	copyright = {2023 The Author(s)},
	issn = {2041-1723},
	url = {https://www.nature.com/articles/s41467-023-36750-3},
	doi = {10.1038/s41467-023-36750-3},
	language = {en},
	number = {1},
	urldate = {2023-04-24},
	journal = {Nature Communications},
	author = {Li, Xiaokang and Machida, Yo and Subedi, Alaska and Zhu, Zengwei and Li, Liang and Behnia, Kamran},
	month = feb,
	year = {2023},
	note = {Number: 1
Publisher: Nature Publishing Group},
	pages = {1027},
}

@article{zhang_gate-tunable_2023,
	title = {Gate-{Tunable} {Phonon} {Magnetic} {Moment} in {Bilayer} {Graphene}},
	volume = {130},
	url = {https://link.aps.org/doi/10.1103/PhysRevLett.130.226302},
	doi = {10.1103/PhysRevLett.130.226302},
	number = {22},
	urldate = {2023-07-20},
	journal = {Physical Review Letters},
	author = {Zhang, Xiao-Wei and Ren, Yafei and Wang, Chong and Cao, Ting and Xiao, Di},
	month = jun,
	year = {2023},
	note = {Publisher: American Physical Society},
	pages = {226302},
}

@article{trifunovic_geometric_2019,
	title = {Geometric orbital magnetization in adiabatic processes},
	volume = {100},
	url = {https://link.aps.org/doi/10.1103/PhysRevB.100.054408},
	doi = {10.1103/PhysRevB.100.054408},
	number = {5},
	urldate = {2023-07-20},
	journal = {Physical Review B},
	author = {Trifunovic, Luka and Ono, Seishiro and Watanabe, Haruki},
	month = aug,
	year = {2019},
	note = {Publisher: American Physical Society},
	pages = {054408},
}

@article{ren_phonon_2021,
	title = {Phonon {Magnetic} {Moment} from {Electronic} {Topological} {Magnetization}},
	volume = {127},
	url = {https://link.aps.org/doi/10.1103/PhysRevLett.127.186403},
	doi = {10.1103/PhysRevLett.127.186403},
	number = {18},
	urldate = {2023-07-20},
	journal = {Physical Review Letters},
	author = {Ren, Yafei and Xiao, Cong and Saparov, Daniyar and Niu, Qian},
	month = oct,
	year = {2021},
	note = {Publisher: American Physical Society},
	pages = {186403},
}

@article{juraschek_dynamical_2017,
	title = {Dynamical multiferroicity},
	volume = {1},
	url = {https://link.aps.org/doi/10.1103/PhysRevMaterials.1.014401},
	doi = {10.1103/PhysRevMaterials.1.014401},
	number = {1},
	urldate = {2023-07-21},
	journal = {Physical Review Materials},
	author = {Juraschek, Dominik M. and Fechner, Michael and Balatsky, Alexander V. and Spaldin, Nicola A.},
	month = jun,
	year = {2017},
	note = {Publisher: American Physical Society},
	pages = {014401},
}

@article{juraschek_orbital_2019,
	title = {Orbital magnetic moments of phonons},
	volume = {3},
	url = {https://link.aps.org/doi/10.1103/PhysRevMaterials.3.064405},
	doi = {10.1103/PhysRevMaterials.3.064405},
	number = {6},
	urldate = {2023-07-21},
	journal = {Physical Review Materials},
	author = {Juraschek, Dominik M. and Spaldin, Nicola A.},
	month = jun,
	year = {2019},
	note = {Publisher: American Physical Society},
	pages = {064405},
}

@article{saparov_lattice_2022,
	title = {Lattice dynamics with molecular {Berry} curvature: {Chiral} optical phonons},
	volume = {105},
	shorttitle = {Lattice dynamics with molecular {Berry} curvature},
	url = {https://link.aps.org/doi/10.1103/PhysRevB.105.064303},
	doi = {10.1103/PhysRevB.105.064303},
	number = {6},
	urldate = {2023-08-19},
	journal = {Physical Review B},
	author = {Saparov, Daniyar and Xiong, Bangguo and Ren, Yafei and Niu, Qian},
	month = feb,
	year = {2022},
	note = {Publisher: American Physical Society},
	pages = {064303},
}

@article{juraschek_giant_2022,
	title = {Giant effective magnetic fields from optically driven chiral phonons in 4f paramagnets},
	volume = {4},
	url = {https://link.aps.org/doi/10.1103/PhysRevResearch.4.013129},
	doi = {10.1103/PhysRevResearch.4.013129},
	number = {1},
	urldate = {2023-08-31},
	journal = {Physical Review Research},
	author = {Juraschek, Dominik M. and Neuman, Tomáš and Narang, Prineha},
	month = feb,
	year = {2022},
	note = {Publisher: American Physical Society},
	pages = {013129},
}

@misc{noauthor_see_nodate,
	title = {See {Supplemental} {Material} at [\{{URL}\} will be inserted by publisher] for derivations, numerical details, and supplemental numerical results, which includes {Refs}. [69- 74].},
}

@article{bistoni_intrinsic_2021,
	title = {Intrinsic {Vibrational} {Angular} {Momentum} from {Nonadiabatic} {Effects} in {Noncollinear} {Magnetic} {Molecules}},
	volume = {126},
	url = {https://link.aps.org/doi/10.1103/PhysRevLett.126.225703},
	doi = {10.1103/PhysRevLett.126.225703},
	number = {22},
	urldate = {2023-09-27},
	journal = {Physical Review Letters},
	author = {Bistoni, Oliviero and Mauri, Francesco and Calandra, Matteo},
	month = jun,
	year = {2021},
	note = {Publisher: American Physical Society},
	pages = {225703},
}

@article{liu_circular_2017,
	title = {Circular {Phonon} {Dichroism} in {Weyl} {Semimetals}},
	volume = {119},
	url = {https://link.aps.org/doi/10.1103/PhysRevLett.119.075301},
	doi = {10.1103/PhysRevLett.119.075301},
	number = {7},
	urldate = {2023-10-27},
	journal = {Physical Review Letters},
	author = {Liu, Donghao and Shi, Junren},
	month = aug,
	year = {2017},
	note = {Publisher: American Physical Society},
	pages = {075301},
}

@article{zabalo_rotational_2022,
	title = {Rotational g factors and {Lorentz} forces of molecules and solids from density functional perturbation theory},
	volume = {105},
	url = {https://link.aps.org/doi/10.1103/PhysRevB.105.094305},
	doi = {10.1103/PhysRevB.105.094305},
	number = {9},
	urldate = {2023-10-31},
	journal = {Physical Review B},
	author = {Zabalo, Asier and Dreyer, Cyrus E. and Stengel, Massimiliano},
	month = mar,
	year = {2022},
	note = {Publisher: American Physical Society},
	pages = {094305},
}

@article{xiao_adiabatically_2021,
	title = {Adiabatically induced orbital magnetization},
	volume = {103},
	url = {https://link.aps.org/doi/10.1103/PhysRevB.103.115432},
	doi = {10.1103/PhysRevB.103.115432},
	number = {11},
	urldate = {2024-11-11},
	journal = {Physical Review B},
	author = {Xiao, Cong and Ren, Yafei and Xiong, Bangguo},
	month = mar,
	year = {2021},
	note = {Publisher: American Physical Society},
	pages = {115432},
}

@article{cheng_large_2020,
	title = {A {Large} {Effective} {Phonon} {Magnetic} {Moment} in a {Dirac} {Semimetal}},
	volume = {20},
	issn = {1530-6984},
	url = {https://doi.org/10.1021/acs.nanolett.0c01983},
	doi = {10.1021/acs.nanolett.0c01983},
	number = {8},
	urldate = {2024-12-04},
	journal = {Nano Letters},
	author = {Cheng, Bing and Schumann, T. and Wang, Youcheng and Zhang, X. and Barbalas, D. and Stemmer, S. and Armitage, N. P.},
	month = aug,
	year = {2020},
	note = {Publisher: American Chemical Society},
	pages = {5991--5996},
}

@article{hernandez_observation_2023,
	title = {Observation of interplay between phonon chirality and electronic band topology},
	volume = {9},
	url = {https://www.science.org/doi/full/10.1126/sciadv.adj4074},
	doi = {10.1126/sciadv.adj4074},
	number = {50},
	urldate = {2025-01-12},
	journal = {Science Advances},
	author = {Hernandez, Felix G. G. and Baydin, Andrey and Chaudhary, Swati and Tay, Fuyang and Katayama, Ikufumi and Takeda, Jun and Nojiri, Hiroyuki and Okazaki, Anderson K. and Rappl, Paulo H. O. and Abramof, Eduardo and Rodriguez-Vega, Martin and Fiete, Gregory A. and Kono, Junichiro},
	month = dec,
	year = {2023},
	note = {Publisher: American Association for the Advancement of Science},
	pages = {eadj4074},
}

@article{ataei_phonon_2024,
	title = {Phonon chirality from impurity scattering in the antiferromagnetic phase of {Sr2IrO4}},
	volume = {20},
	copyright = {2024 The Author(s), under exclusive licence to Springer Nature Limited},
	issn = {1745-2481},
	url = {https://www.nature.com/articles/s41567-024-02384-5},
	doi = {10.1038/s41567-024-02384-5},
	language = {en},
	number = {4},
	urldate = {2025-02-23},
	journal = {Nature Physics},
	author = {Ataei, A. and Grissonnanche, G. and Boulanger, M.-E. and Chen, L. and Lefrançois, É and Brouet, V. and Taillefer, L.},
	month = apr,
	year = {2024},
	note = {Publisher: Nature Publishing Group},
	pages = {585--588},
}

@article{shabala_phonon_2024,
	title = {Phonon {Inverse} {Faraday} {Effect} from {Electron}-{Phonon} {Coupling}},
	volume = {133},
	url = {https://link.aps.org/doi/10.1103/PhysRevLett.133.266702},
	doi = {10.1103/PhysRevLett.133.266702},
	number = {26},
	urldate = {2025-03-05},
	journal = {Physical Review Letters},
	author = {Shabala, Natalia and Geilhufe, R. Matthias},
	month = dec,
	year = {2024},
	note = {Publisher: American Physical Society},
	pages = {266702},
}

@article{baydin_magnetic_2022,
	title = {Magnetic {Control} of {Soft} {Chiral} {Phonons} in {PbTe}},
	volume = {128},
	url = {https://link.aps.org/doi/10.1103/PhysRevLett.128.075901},
	doi = {10.1103/PhysRevLett.128.075901},
	number = {7},
	urldate = {2025-04-21},
	journal = {Physical Review Letters},
	author = {Baydin, Andrey and Hernandez, Felix G. G. and Rodriguez-Vega, Martin and Okazaki, Anderson K. and Tay, Fuyang and Noe, G. Timothy and Katayama, Ikufumi and Takeda, Jun and Nojiri, Hiroyuki and Rappl, Paulo H. O. and Abramof, Eduardo and Fiete, Gregory A. and Kono, Junichiro},
	month = feb,
	year = {2022},
	note = {Publisher: American Physical Society},
	pages = {075901},
}

@article{klebl_ultrafast_2025,
	title = {Ultrafast {Pseudomagnetic} {Fields} from {Electron}-{Nuclear} {Quantum} {Geometry}},
	volume = {134},
	url = {https://link.aps.org/doi/10.1103/PhysRevLett.134.016705},
	doi = {10.1103/PhysRevLett.134.016705},
	number = {1},
	urldate = {2025-04-29},
	journal = {Physical Review Letters},
	author = {Klebl, Lennart and Schobert, Arne and Eckstein, Martin and Sangiovanni, Giorgio and Balatsky, Alexander V. and Wehling, Tim O.},
	month = jan,
	year = {2025},
	note = {Publisher: American Physical Society},
	pages = {016705},
}

@misc{chen_geometric_2025,
	title = {Geometric {Origin} of {Phonon} {Magnetic} {Moment} in {Dirac} {Materials}},
	url = {http://arxiv.org/abs/2505.09732},
	doi = {10.48550/arXiv.2505.09732},
	urldate = {2025-05-16},
	publisher = {arXiv},
	author = {Chen, Wenqin and Zhang, Xiao-Wei and Cao, Ting and Lin, Shi-Zeng and Xiao, Di},
	month = may,
	year = {2025},
	note = {arXiv:2505.09732 [cond-mat]},
}

@article{chen_gauge_2025,
	title = {Gauge theory of giant phonon magnetic moment in doped {Dirac} semimetals},
	volume = {111},
	url = {https://link.aps.org/doi/10.1103/PhysRevB.111.035126},
	doi = {10.1103/PhysRevB.111.035126},
	number = {3},
	urldate = {2025-05-27},
	journal = {Physical Review B},
	author = {Chen, Wenqin and Zhang, Xiao-Wei and Su, Ying and Cao, Ting and Xiao, Di and Lin, Shi-Zeng},
	month = jan,
	year = {2025},
	note = {Publisher: American Physical Society},
	pages = {035126},
}

@article{che_magnetic_2025,
	title = {Magnetic {Order} {Induced} {Chiral} {Phonons} in a {Ferromagnetic} {Weyl} {Semimetal}},
	volume = {134},
	url = {https://link.aps.org/doi/10.1103/PhysRevLett.134.196906},
	doi = {10.1103/PhysRevLett.134.196906},
	number = {19},
	urldate = {2025-05-28},
	journal = {Physical Review Letters},
	author = {Che, Mengqian and Liang, Jinxuan and Cui, Yunpeng and Li, Hao and Lu, Bingru and Sang, Wenbo and Li, Xiang and Dong, Xuebin and Zhao, Le and Zhang, Shuai and Sun, Tao and Jiang, Wanjun and Liu, Enke and Jin, Feng and Zhang, Tiantian and Yang, Luyi},
	month = may,
	year = {2025},
	note = {Publisher: American Physical Society},
	pages = {196906},
}

@article{yang_inherent_2025,
	title = {Inherent {Circular} {Dichroism} of {Phonons} in {Magnetic} {Weyl} {Semimetal} \$\{{\textbackslash}mathrm\{{Co}\}\}\_\{3\}\{{\textbackslash}text\{{Sn}\}\}\_\{2\}\{{\textbackslash}mathrm\{{S}\}\}\_\{2\}\$},
	volume = {134},
	url = {https://link.aps.org/doi/10.1103/PhysRevLett.134.196905},
	doi = {10.1103/PhysRevLett.134.196905},
	number = {19},
	urldate = {2025-05-28},
	journal = {Physical Review Letters},
	author = {Yang, R. and Zhu, Y.-Y. and Steigleder, M. and Liu, Y.-C. and Liu, C.-C. and Qiu, X.-G. and Zhang, Tiantian and Dressel, M.},
	month = may,
	year = {2025},
	note = {Publisher: American Physical Society},
	pages = {196905},
}

@article{zhu_theory_2012,
	title = {Theory of orbital magnetization in disordered systems},
	volume = {86},
	copyright = {http://link.aps.org/licenses/aps-default-license},
	issn = {1098-0121, 1550-235X},
	url = {https://link.aps.org/doi/10.1103/PhysRevB.86.214415},
	doi = {10.1103/PhysRevB.86.214415},
	language = {en},
	number = {21},
	urldate = {2025-06-19},
	journal = {Physical Review B},
	author = {Zhu, Guobao and Yang, Shengyuan A. and Fang, Cheng and Liu, W. M. and Yao, Yugui},
	month = dec,
	year = {2012},
	pages = {214415},
}

@article{onoda_theory_2006,
	title = {Theory of {Non}-{Equilibirum} {States} {Driven} by {Constant} {Electromagnetic} {Fields}: — {Non}-{Commutative} {Quantum} {Mechanics} in the {Keldysh} {Formalism} —},
	volume = {116},
	issn = {0033-068X},
	shorttitle = {Theory of {Non}-{Equilibirum} {States} {Driven} by {Constant} {Electromagnetic} {Fields}},
	url = {https://doi.org/10.1143/PTP.116.61},
	doi = {10.1143/PTP.116.61},
	number = {1},
	urldate = {2025-06-19},
	journal = {Progress of Theoretical Physics},
	author = {Onoda, Shigeki and Sugimoto, Naoyuki and Nagaosa, Naoto},
	month = jul,
	year = {2006},
	pages = {61--86},
}

@article{oh_phonon_2025,
	title = {Phonon {Thermal} {Hall} {Effect} in {Mott} {Insulators} via {Skew} {Scattering} by the {Scalar} {Spin} {Chirality}},
	volume = {15},
	url = {https://link.aps.org/doi/10.1103/PhysRevX.15.011036},
	doi = {10.1103/PhysRevX.15.011036},
	number = {1},
	urldate = {2025-07-12},
	journal = {Physical Review X},
	author = {Oh, Taekoo and Nagaosa, Naoto},
	month = feb,
	year = {2025},
	note = {Publisher: American Physical Society},
	pages = {011036},
}

@misc{shitade_gradient_2017,
	title = {Gradient expansion formalism for generic spin torques},
	url = {http://arxiv.org/abs/1708.03424},
	doi = {10.48550/arXiv.1708.03424},
	urldate = {2025-07-15},
	publisher = {arXiv},
	author = {Shitade, Atsuo},
	month = aug,
	year = {2017},
	note = {arXiv:1708.03424 [cond-mat]},
}

@article{ismail-beigi_coupling_2001,
	title = {Coupling of {Nonlocal} {Potentials} to {Electromagnetic} {Fields}},
	volume = {87},
	url = {https://link.aps.org/doi/10.1103/PhysRevLett.87.087402},
	doi = {10.1103/PhysRevLett.87.087402},
	number = {8},
	urldate = {2025-07-23},
	journal = {Physical Review Letters},
	author = {Ismail-Beigi, Sohrab and Chang, Eric K. and Louie, Steven G.},
	month = aug,
	year = {2001},
	note = {Publisher: American Physical Society},
	pages = {087402},
}

@article{chen_unified_2011,
	title = {Unified formalism for calculating polarization, magnetization, and more in a periodic insulator},
	volume = {84},
	url = {https://link.aps.org/doi/10.1103/PhysRevB.84.205137},
	doi = {10.1103/PhysRevB.84.205137},
	number = {20},
	urldate = {2025-08-01},
	journal = {Physical Review B},
	author = {Chen, Kuang-Ting and Lee, Patrick A.},
	month = nov,
	year = {2011},
	note = {Publisher: American Physical Society},
	pages = {205137},
}

@book{giuliani_quantum_2005,
	edition = {1},
	title = {Quantum {Theory} of the {Electron} {Liquid}},
	copyright = {https://www.cambridge.org/core/terms},
	isbn = {978-0-521-52796-5 978-0-521-82112-4 978-0-511-61991-5},
	url = {https://www.cambridge.org/core/product/identifier/9780511619915/type/book},
	urldate = {2025-08-06},
	publisher = {Cambridge University Press},
	author = {Giuliani, Gabriele and Vignale, Giovanni},
	month = mar,
	year = {2005},
	doi = {10.1017/CBO9780511619915},
}

@misc{hartnoll_holographic_2018,
	title = {Holographic quantum matter},
	url = {http://arxiv.org/abs/1612.07324},
	doi = {10.48550/arXiv.1612.07324},
	urldate = {2025-08-27},
	publisher = {arXiv},
	author = {Hartnoll, Sean A. and Lucas, Andrew and Sachdev, Subir},
	month = mar,
	year = {2018},
	note = {arXiv:1612.07324 [hep-th]},
}

@article{lau_topological_2019,
	title = {Topological {Semimetals} in the {SnTe} {Material} {Class}: {Nodal} {Lines} and {Weyl} {Points}},
	volume = {122},
	issn = {0031-9007, 1079-7114},
	shorttitle = {Topological {Semimetals} in the {SnTe} {Material} {Class}},
	url = {https://link.aps.org/doi/10.1103/PhysRevLett.122.186801},
	doi = {10.1103/PhysRevLett.122.186801},
	language = {en},
	number = {18},
	urldate = {2025-09-08},
	journal = {Physical Review Letters},
	author = {Lau, Alexander and Ortix, Carmine},
	month = may,
	year = {2019},
	pages = {186801},
}

@article{xue_extrinsic_2025,
	title = {Extrinsic {Mechanisms} of {Phonon} {Magnetic} {Moment}},
	volume = {135},
	url = {https://link.aps.org/doi/10.1103/6m7v-p99w},
	doi = {10.1103/6m7v-p99w},
	number = {10},
	urldate = {2025-11-06},
	journal = {Physical Review Letters},
	author = {Xue, Rui and Qiao, Zhenhua and Gao, Yang and Niu, Qian},
	month = sep,
	year = {2025},
	note = {Publisher: American Physical Society},
	pages = {106605},
}

@misc{shabala_axial_2025,
	title = {Axial phono-magnetic effects},
	url = {http://arxiv.org/abs/2511.03329},
	doi = {10.48550/arXiv.2511.03329},
	urldate = {2025-12-17},
	publisher = {arXiv},
	author = {Shabala, Natalia and Tietjen, Finja and Geilhufe, R. Matthias},
	month = nov,
	year = {2025},
	note = {arXiv:2511.03329 [cond-mat]},
}

@article{mangeolle_extrinsic_2026,
	title = {Extrinsic {Contribution} to {Bosonic} {Thermal} {Hall} {Transport}},
	volume = {16},
	url = {https://link.aps.org/doi/10.1103/grzx-v6sj},
	doi = {10.1103/grzx-v6sj},
	number = {1},
	urldate = {2026-03-06},
	journal = {Physical Review X},
	author = {Mangeolle, Léo and Knolle, Johannes},
	month = mar,
	year = {2026},
	note = {Publisher: American Physical Society},
	pages = {011048},
}

@article{delaire_giant_2011,
	title = {Giant anharmonic phonon scattering in {PbTe}},
	volume = {10},
	copyright = {2011 Springer Nature Limited},
	issn = {1476-4660},
	url = {https://www.nature.com/articles/nmat3035},
	doi = {10.1038/nmat3035},
	language = {en},
	number = {8},
	urldate = {2026-03-06},
	journal = {Nature Materials},
	author = {Delaire, O. and Ma, J. and Marty, K. and May, A. F. and McGuire, M. A. and Du, M.-H. and Singh, D. J. and Podlesnyak, A. and Ehlers, G. and Lumsden, M. D. and Sales, B. C.},
	month = aug,
	year = {2011},
	note = {Publisher: Nature Publishing Group},
	pages = {614--619},
}

@article{ribeiro_strong_2018,
	title = {Strong anharmonicity in the phonon spectra of {PbTe} and {SnTe} from first principles},
	volume = {97},
	issn = {2469-9950, 2469-9969},
	url = {https://link.aps.org/doi/10.1103/PhysRevB.97.014306},
	doi = {10.1103/PhysRevB.97.014306},
	language = {en},
	number = {1},
	urldate = {2026-03-10},
	journal = {Physical Review B},
	author = {Ribeiro, Guilherme A. S. and Paulatto, Lorenzo and Bianco, Raffaello and Errea, Ion and Mauri, Francesco and Calandra, Matteo},
	month = jan,
	year = {2018},
	pages = {014306},
}

@article{assaf_magnetooptical_2017,
	title = {Magnetooptical determination of a topological index},
	volume = {2},
	issn = {2397-4648},
	url = {https://www.nature.com/articles/s41535-017-0028-5},
	doi = {10.1038/s41535-017-0028-5},
	language = {en},
	number = {1},
	urldate = {2026-03-13},
	journal = {npj Quantum Materials},
	author = {Assaf, Badih A. and Phuphachong, Thanyanan and Volobuev, Valentine V. and Bauer, Günther and Springholz, Gunther and De Vaulchier, Louis-Anne and Guldner, Yves},
	month = may,
	year = {2017},
	pages = {26},
}

@article{juraschek_chiral_2025,
	title = {Chiral phonons},
	volume = {21},
	issn = {1745-2473, 1745-2481},
	url = {https://www.nature.com/articles/s41567-025-03001-9},
	doi = {10.1038/s41567-025-03001-9},
	language = {en},
	number = {10},
	urldate = {2026-03-14},
	journal = {Nature Physics},
	author = {Juraschek, Dominik M. and Geilhufe, R. Matthias and Zhu, Hanyu and Basini, Martina and Baum, Peter and Baydin, Andrey and Chaudhary, Swati and Fechner, Michael and Flebus, Benedetta and Grissonnanche, Gael and Kirilyuk, Andrei I. and Lemeshko, Mikhail and Maehrlein, Sebastian F. and Mignolet, Maxime and Murakami, Shuichi and Niu, Qian and Nowak, Ulrich and Romao, Carl P. and Rostami, Habib and Satoh, Takuya and Spaldin, Nicola A. and Ueda, Hiroki and Zhang, Lifa},
	month = oct,
	year = {2025},
	pages = {1532--1540},
}

@article{zhu_observation_2018,
	title = {Observation of chiral phonons},
	volume = {359},
	issn = {0036-8075, 1095-9203},
	url = {https://www.science.org/doi/10.1126/science.aar2711},
	doi = {10.1126/science.aar2711},
	language = {en},
	number = {6375},
	urldate = {2026-03-14},
	journal = {Science},
	author = {Zhu, Hanyu and Yi, Jun and Li, Ming-Yang and Xiao, Jun and Zhang, Lifa and Yang, Chih-Wen and Kaindl, Robert A. and Li, Lain-Jong and Wang, Yuan and Zhang, Xiang},
	month = feb,
	year = {2018},
	pages = {579--582},
}

@article{zhang_angular_2014,
	title = {Angular {Momentum} of {Phonons} and the {Einstein}--de {Haas} {Effect}},
	volume = {112},
	url = {https://link.aps.org/doi/10.1103/PhysRevLett.112.085503},
	doi = {10.1103/PhysRevLett.112.085503},
	number = {8},
	urldate = {2026-03-14},
	journal = {Physical Review Letters},
	author = {Zhang, Lifa and Niu, Qian},
	month = feb,
	year = {2014},
	note = {Publisher: American Physical Society},
	pages = {085503},
}

@article{xu_nondegenerate_2018,
	title = {Nondegenerate chiral phonons in the {Brillouin}-zone center of \${\textbackslash}sqrt\{3\}{\textbackslash}ifmmode{\textbackslash}times{\textbackslash}else{\textbackslash}texttimes{\textbackslash}fi\{\}{\textbackslash}sqrt\{3\}\$ honeycomb superlattices},
	volume = {98},
	url = {https://link.aps.org/doi/10.1103/PhysRevB.98.134304},
	doi = {10.1103/PhysRevB.98.134304},
	number = {13},
	urldate = {2026-03-14},
	journal = {Physical Review B},
	author = {Xu, Xifang and Chen, Hao and Zhang, Lifa},
	month = oct,
	year = {2018},
	note = {Publisher: American Physical Society},
	pages = {134304},
}

@article{sasaki_magnetization_2021,
	title = {Magnetization control by angular momentum transfer from surface acoustic wave to ferromagnetic spin moments},
	volume = {12},
	copyright = {2021 The Author(s)},
	issn = {2041-1723},
	url = {https://www.nature.com/articles/s41467-021-22728-6},
	doi = {10.1038/s41467-021-22728-6},
	language = {en},
	number = {1},
	urldate = {2026-03-14},
	journal = {Nature Communications},
	author = {Sasaki, R. and Nii, Y. and Onose, Y.},
	month = may,
	year = {2021},
	note = {Publisher: Nature Publishing Group},
	pages = {2599},
}

@article{nova_effective_2017,
	title = {An effective magnetic field from optically driven phonons},
	volume = {13},
	copyright = {2016 Springer Nature Limited},
	issn = {1745-2481},
	url = {https://www.nature.com/articles/nphys3925},
	doi = {10.1038/nphys3925},
	language = {en},
	number = {2},
	urldate = {2026-03-14},
	journal = {Nature Physics},
	author = {Nova, T. F. and Cartella, A. and Cantaluppi, A. and Först, M. and Bossini, D. and Mikhaylovskiy, R. V. and Kimel, A. V. and Merlin, R. and Cavalleri, A.},
	month = feb,
	year = {2017},
	note = {Publisher: Nature Publishing Group},
	pages = {132--136},
}

@misc{li_phonon_2025,
	title = {Phonon {Dichroisms} {Revealing} {Unusual} {Electronic} {Quantum} {Geometry}},
	url = {http://arxiv.org/abs/2511.16141},
	doi = {10.48550/arXiv.2511.16141},
	urldate = {2026-03-16},
	publisher = {arXiv},
	author = {Li, Ding and Yang, Guoao and Qin, Tao and Zhou, Jianhui and Yao, Yugui},
	month = nov,
	year = {2025},
	note = {arXiv:2511.16141 [cond-mat]},
}

\end{document}